\documentclass{sigchi}

\setcopyright{rightsretained}
\doi{}
\isbn{}

\usepackage{balance}       
\usepackage{graphics}      
\usepackage[T1]{fontenc}   
\usepackage{mathpazo}
\usepackage{txfonts}
\usepackage{mathptmx}
\usepackage[pdflang={en-US},pdftex]{hyperref}
\usepackage{color}
\usepackage{booktabs}
\usepackage{textcomp}
\usepackage{microtype}        
\usepackage{ccicons}          

\usepackage{todonotes}

\def\plaintitle{SIGCHI Conference Proceedings Format}

\def\emptyauthor{}
\def\plainkeywords{Personalization;animation;emotion;engagement; empathy; self-reflection.}

\makeatletter
\def\url@leostyle{%
  \@ifundefined{selectfont}{
    \def\UrlFont{\sf}
  }{
    \def\UrlFont{\small\bf\ttfamily}
  }}
\makeatother
\def\pprw{8.5in}
\def\pprh{11in}

\definecolor{linkColor}{RGB}{6,125,233}
\hypersetup{%
  pdftitle={\plaintitle},
  pdfauthor={\emptyauthor},
  pdfkeywords={\plainkeywords},
  pdfdisplaydoctitle=true, 
  bookmarksnumbered,
  pdfstartview={FitH},
  colorlinks,
  citecolor=black,
  filecolor=black,
  linkcolor=black,
  urlcolor=linkColor,
  breaklinks=true,
  hypertexnames=false
}

\begin{document}

\title{A Trip to the Moon: Personalized Animated Movies for Self-reflection}

\numberofauthors{4}
\author{%
\alignauthor Fengjiao Peng\\
       \affaddr{MIT Media Lab}\\
       \affaddr{Cambridge, MA, USA}\\
       \email{fpeng@mit.edu}
\alignauthor Veronica LaBelle\\
       \affaddr{MIT}\\
       \affaddr{Cambridge, MA, USA}\\
       \email{vlabelle@mit.edu}
\alignauthor Emily Yue\\
       \affaddr{Harvard University}\\
       \affaddr{Cambridge, MA, USA}\\
       \email{eyue@college.harvard.edu}
\alignauthor Rosalind Picard\\
       \affaddr{MIT Media Lab}\\
       \affaddr{Cambridge, MA, USA}\\
       \email{picard@media.mit.edu}
}
\maketitle

\begin{abstract}
Self-tracking physiological and psychological data poses the challenge of presentation and interpretation. Insightful narratives for self-tracking data can motivate the user towards constructive self-reflection. One powerful form of narrative that engages audience across various culture and age groups is animated movies. We collected a week of self-reported mood and behavior data from each user and created in Unity a personalized animation based on their data. We evaluated the impact of their video in a randomized control trial with a non-personalized animated video as control. We found that personalized videos tend to be more emotionally engaging, encouraging greater and lengthier writing that indicated self-reflection about moods and behaviors, compared to non-personalized control videos.
\end{abstract}

\category{H.5.1.}
{Information interfaces and presentation (e.g., HCI): Animations.}{}{}

\keywords{Personalization; animation; emotion; engagement; empathy; self-reflection.}

\section{Introduction}

The development of mobile phone technology and biological sensors is enabling individuals to self-track biological, physical and environmental information. From rich self-tracking data, individuals can interpret and infer the patterns, correlations and causal relations in their own behavior and wellbeing. Recent HCI research attempts to help users build up a critical understanding of their data, especially to motivate users to change their behaviors for the betterment of their own wellbeing or their social, biological environment \cite{sengers2005reflective}. We often need to couple such critical thinking with emotional engagement, as emotions are a powerful factor in both impairing and facilitating decision making \cite{picard1997affective}.

Storytelling or narratives are a fundamental way in which human beings make sense of the world \cite{mateas2003narrative}. To help users interpret self-tracking data, we are often trying to retell the story behind the data. To construct the appropriate narrative for different reflective purposes, HCI designers deploy a wide range of design languages. The language could be quantitative, iterating through the data or summarizing them in statistics, tables, or graphs. Others choose a more qualitative, implicit approach, embedding meanings in dynamic visuals, audio, and physical forms, allowing for less precise and more ambiguous feedback to the user. (See H\"{o}\"{o}k's \textit{AffectCam} \cite{sas2013affectcam} and \textit{Affective Diary} \cite{lindstrom2006affective}) Sengers points out that ambiguity and multiple interpretations in HCI systems can often open up new design spaces that converge with art and humanities \cite{sengers2006staying}. 

We've seen traditional quantitative reflective designs made powerful with an intelligent narrative, such as in Hans Rosling's talks on the visualization of statistics \cite{rosling2006debunking}. The statistics are brought to life by a dynamic, emotional narrative, which induces audience's curiosity and empathy towards the story behind the figures. It takes storytelling skills and a deep understanding of the background knowledge to create these engaging narratives, which is difficult to realize for chaotic and often sparse self-tracking data on an individual basis, where the causal and correlational relations are hard to capture. Furthermore, an engaging narrative for personal data should be based on a knowledge of personal context. For example, when a mobile phone app records reduced exercise rate, it could be either due to a downturn of mood or a vacation. The narrative to motivate the user to get back on track in exercise should take into account these personal contextual differences.

We attempt to create emotional, personal narratives of data related to behavior and mental wellbeing by making animated movies customized to each individual's emotional trajectory. Readers can compare our story world to the animated movie \textit{Inside Out}, where emotions are symbolized as characters and environment designs. To make it practical for a large number of users, we created a novel fully automated animation generator that works by parameterizing animations, within a game engine. The parameters are derived from each user's personal data stream that they input over a week. We thus deploy a design language for "seeing their data" that most users find familiar and easy to engage with emotionally: the filmic language of animation movies. 

HCI and conversational agent designers have successfully transplanted animation principles to communicate affective states both from users and to users across different cultural backgrounds. \cite{isbister2007sensual,bates1994role,van2004bringing}  These principles are often referred to as "illusions of life", designed to bring lifelike qualities to animated characters \cite{porter2000site}. The core notion is that characters' actions should reflect intelligence, emotion and personality \cite{thomas1995illusion}. In animated movies, these animation principles, as well as a cohesive storyline, dedicated cinematography, light, and atmospheric and sound design work together to engage the viewers in the characters' twisting and turning emotional trajectory.

Our animated movies are personalized on two levels: explicit and implicit. On the explicit level, the cinematographic art and the storyline vary according to the input data of mood and behavior across time. On the implicit level, we do not try to enact the user's past experience precisely through animation. The scenes aim at creating general mood and atmosphere, indicating instead of elaborating on the often complex and tangled emotions experienced by the user. The Barnum effect states the tendency for individuals to accept general and ambiguous personality tests outcome descriptions as accurate if they are told that the results are tailored for them. \cite{snyder1977acceptance} We hypothesize that by knowing the animated movie is customized to their data, users will see more than is visualized in the movie and relate implicit animations and filmic effects to explicit past experiences. In our experiment, 17 out of 23 participants were able to connect the movie with their past experience when asked, "what do you think the video reflected about you".

While physiological, behavioral and self-report data can be used to track user's psychological processes \cite{picard2005evaluating}, in this work, we use self-reported mood and behavior data for three reasons: First, we can avoid challenges in the interpretation of physiology data, such as the difficulty in distinguishing between joy and anger \cite{bentley2005evaluation}; Second, users can recall and implicitly select the affective states they wish to have feedback on, which promotes mindfulness and awareness of their emotions \cite{baer2006using}; Third, users are required to recall the past 24 hours every day, so their memory of the events should be relatively reliable \cite{kihlstrom2000emotion}.

The goal of this project is to open up the design space where users' mood and behavior data are relived in an emotional narrative, in order to maximize viewer's self-compassion and motivation to maintain or change their behavior post viewing. We demonstrate in a one-week randomized control study that while both animation don't involve a traditional "story" narrative, users are better at interpreting personalized animations compared to a generic animation; that users are able to connect to the animated agent, engage with the animated movies emotionally and constructively reflect on their mood and behavior patterns. 

\begin{figure}
\centering
  \includegraphics[width=0.5\columnwidth]{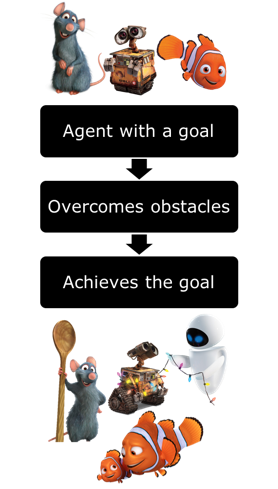}
  \caption{A common story structure for animated feature films. The rat, robot and clown fish characters are from Ratatouille (Pixar, 2007), Wall-E (Pixar, 2008) and Finding Nemo (Pixar, 2003).}
  \label{fig:story_structure}
\end{figure}

\section{Related Work}
This section summarizes previous research on designing HCI systems to reflect affective states and building motivational narratives for self-reflection and behavior change.

\subsection{Reflective Design}
Reflective design focuses on embedding unconscious cultural assumptions and values in the design of IT systems, in particular to promote a user's critical understanding of such values \cite{sengers2005reflective}. H\"{o}\"{o}k proposed the concept of the affective loop experience where the users are pulled into either virtual or bodily feedback loops to their own emotions \cite{hook2009affective}. 

Implicit reflective systems can involve both quantitative and qualitative design languages. \textit{EmotionCheck} is an emotion-regulation technology featuring a wearable device designed to provide a false, lower heart rate to the users to regulate their anxiety level \cite{costa2016emotioncheck}. The regulation is done without users being implicitly involved in trying to reduce their heart rate. Other approaches use visual or physical forms to embody the signals. \textit{AffectCam} \cite{sas2013affectcam} uses physiological measures of arousal to select the most memorable photos in an individual's ordinary life. \textit{Affective Diary} \cite{lindstrom2006affective} incorporates the bodily experience of movement and arousal in an individual's journaling process. These designs support and sustain the reflection and reliving of past experiences. Users are able to discover correlations between behaviors and the affective reflections which may give new insight and inspire life changes \cite{lindstrom2006affective}. Our work is similar in its efforts to review past experiences, but does so while superimposing an emotional response through the animation effects.

\begin{figure*}[t]
\centering
  \includegraphics[width=\textwidth]{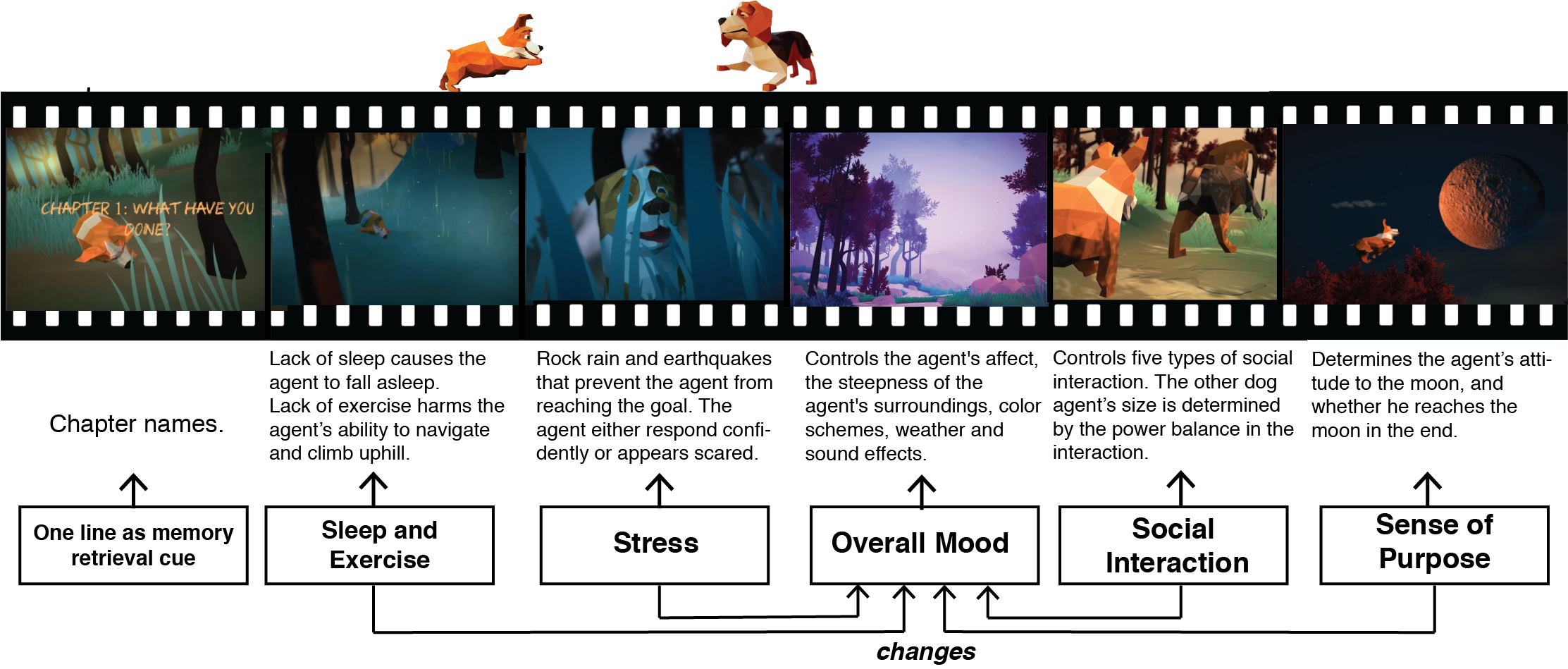}
  \caption{System overview. Six types of data (overall mood, sleep and exercise, social interaction, sense of purpose, stress, and one line as memory retrieval cue) dictate the story.}
  \label{fig:system_overview}
\end{figure*}

\subsection{Motivational Narrative Systems}
Health, education and training can be incorporated in gaming narratives where users are motivated by in-game rewards \cite{michael2005serious}. The study of Roth \textit{et al} \cite{roth2009motivational} explored the factors that give interactive storytelling in video games motivational appeal for the user: curiosity, suspense, aesthetic pleasure and self-enhancement. In the game for health designed by Gobel \textit{et al} \cite{gobel2010serious}, user's body movement, heart rate, pulse and personalized workout data are used to influence the storytelling of the video game. Gustafsson \textit{et al} \cite{gustafsson2009power} designed a mobile-phone based game linked to the power consumption of the house, where teenagers are motivated to keep the virtual monster avatar healthy through lowering the power consumption. Unlike traditional motivational games, our approach focuses on building emotional connection with the agent by a determined, non-interactive story. We focus on the passive consumption of the narrative and the loss of control, which fundamentally distinguish movies and TV from gaming, and their different cognitive and psychological effects on the user.

\section{System Overview}

Several empirical principles in animated film making provide an intuition for creating this work. In Disney and Pixar animated movies, the agent (main character) often has a simple goal and a limited skill set to achieve the goal. The exposition, rising action, climax and falling action in traditional storytelling \cite{freytag1896freytag} all spring from obstacles that prevents the agent from achieving the goal. (Fig. \ref{fig:story_structure}) The goal is common and broad enough to conform to a majority of viewer and life situations. The symbolic form of goal also signifies the flow of time and metamorphosis in the agent. While the obstacles bring depth to the story, the agent can have a simple personality, allowing both children and adults to relate.

Based on the story structure, we create a narrative where a corgi dog (agent) travels and attempts to go to the moon (goal) where he/she either succeeds or fails. We depict the dog's journey by means of it following an ever-changing path, based on the personal data, representative of traveling through life. Along this journey, the dog will face various obstacles, in the form of weather, rocks blocking his way to the moon, and more, of which the nuances and variations will be dictated by the user's self-reported mood and behavior data, recorded every day over a week. The dog will also encounter social interactions (with other dog agents) that can elicit a variety of feelings.

\label{section:data} Six main groups of mood and behavior data are collected from the user through daily surveys: overall mood today, sleep and exercise, stress, social interaction, sense of purpose, and a line that reminds the user of that day. The questions on overall mood, sleep and exercise, stress and social interaction are adapted from Sano's study to measure stress, sleep and happiness, where one's mental wellbeing is systematically evaluated \cite{sano2016measuring}. Sleep cycles, exercise and social interaction are correlated with mood \cite{sano2016measuring}. The questions on sense of purpose are adapted from the purpose-in-life test, as the sense of purpose is highly correlated with depression and other mental health problems \cite{harlow1986depression}. After a preliminary study, we found that it's often difficult for users to recall the events on a day as far as a week ago by just telling them "last Tuesday". Therefore we ask the users for one line as memory retrieval cue, which is visualized as chapter titles in the animation. In the experiment, we received cues on a spectrum, varying from concrete ones, such as "Back to work after Labor Day...", "Rainy beautiful sunset KKC", "Pho Basil", to more abstract ones such as "Idiotic", "defused bombs", "tell me why, yang", "You are beautiful".

\subsection{From Data to Animation}

We built a 3D environment in Unity and used a polygon dog model as the agent. We chose dogs as our avatars because they have intuitive expressions of emotions, with which we can avoid the "uncanny valley" problems of humanoid avatars. The agent's animation is created by experienced animators to display distinct emotions through his/her facial expression and body movement, each story event being the combination of several animations. When the animation starts, the agent navigates to the goal, stopping whenever an event happens. Fig. \ref{fig:system_overview} shows the mapping between data and animation events. 

The choices of animation are made by three researchers based on their prior experience in filming or animation, and computational constraints. For example, stress is represented by rocks falling from the sky. In early iterations, we experimented with a mountain appearing, forming visual obstacle in front of the agent. However, re-computing the agent's navigation path while the landscape was changing raised computational issues, causing the agent to sometimes walk underground. Future work, incorporating varied environments, can add additional visualizations of stressors.

\textbf{One line that will remind you of today.} This line is used as chapter titles in the animation, and each day is a chapter. Due to the sparsity of data (users occasionally fail to fill out daily surveys) and repetitiveness of "normal" days, the four most "dramatic" days of the week, i.e. the data peak and valleys, are used in the animation.

\begin{figure}
\centering
  \includegraphics[width=\columnwidth]{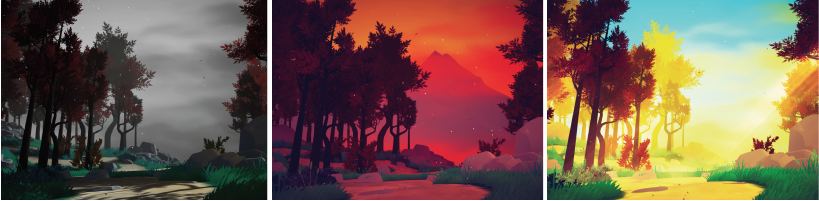}
  \caption{Three color and weather schemes that correspond to depression, anger and excitement.}
  \label{fig:mood}
\end{figure}

\textbf{Overall Mood.} We map each users' overall feeling to a valence and arousal chart, using Likert scales of 1-7 that users would select in the surveys. Different points on the valence-arousal graph are mapped to different states of the agent's body animation and facial expression and moving speed; the steepness of the agent's surroundings, the color and weather embedded in the environment; the forest sound and music effects. For example, when the mood is calm and peaceful, the agent wanders leisurely with a content smile, the ground is smooth, the color scheme is a light purple with a thin layer of smoke, and the forest sounds of bird chirping and breezes. All the factors are designed to express the emotion experienced in the day. (Fig. \ref{fig:mood})

\textbf{Daily events.} Users report social interactions, bed time, sleep length and quality, and amount of exercise every day. The social interaction between the agent and other dog agents depends on how much social interaction there is, what kind of interaction it is and who the user interacted with. General types of interaction include: no interaction, where the corgi acts with loneliness or content solitude, neutral interactions that don't affect mood, a happy, playful interaction, an angry fight, rejection. Depending on who the user interacts with, we introduce a power dynamic through the size of other dog agents that our corgi agent interacts with, where smaller would imply interacting with someone more submissive, inferior or childlike, and larger would imply interacting with someone more dominant or authoritative. 

Similar to the social interactions, sleep and exercise events are scripted to depict users' data in that category. Too little sleep would affect performance in the agent's movement, and eventually cause him or her to fall asleep. A good amount of exercise will make the agent appear more energetic and better overcome the steep landscape.

\textbf{Stress.} \label{section:stress} Stress is another factor that is closely related to mood, and often requires user's active regulation. Facing the same stressor, some view it as a challenge, a source of motivation, while others view it as a threat \cite{mccrae1984situational}. Therefore, we designed separate questions "how stressful is today" and "how was your experience handling the stress." The first question investigates users' perceived level of stress, and the second question investigates how resilient or vulnerable they are when facing the stress. To answer "how was your experience handling the stress," users can select on a Likert scale of 0-7, where 0 corresponds to "I was anxious and stressed out", and 7 corresponds to "I felt things were under control". Stressors are visualized in the movie as a rain of rocks that blocks him from reaching the moon. Depending on the user's data, the agent either appears to be scared by the threat or bravely overcomes it.

\textbf{Sense of purpose and achievement.} We investigate users' sense of purpose using questions adapted from the Purpose-in-life psychology test \cite{harlow1986depression}. How much interest the user has in life, how purposeful he feels, and how much personal achievement on the days investigated decide together the distance from the agent to the moon and whether the agent will have enough momentum to travel to it in the end.

\textbf{Cinematography.} Sound design is incorporated to express the corgi's affective states: panting during a heavy run, snoring while sleeping, howling, barking, and so on.

\begin{table*}[t]
  \centering
  \begin{tabular}{l r r r r r r}
    {\small\textit{Animated Affect}}
    & {\small \textit{Happiness}}
      & {\small \textit{Calmness}}
    & {\small \textit{Is the situation stressful?}}
    & {\small \textit{Energetic}}
     & {\small \textit{Sleep quality}}
      & {\small \textit{Sociable}}
    \\
    \midrule
    a) stressed, anxious & 3.86$\pm$2.06 & 5$\pm$1.98 & 5.33$\pm$1.66 & - & - & - \\
    b) happy, energetic & 6.23$\pm$1.35 & 4.73$\pm$2.02 & 2.33$\pm$1.65 & 6.5$\pm$0.63 & 5.9$\pm$1.58 & 6.23$\pm$2.01 \\
    c) frustrated, fatigue & 1.7$\pm$1.12 & 2.43$\pm$1.48 &- & 1.3$\pm$0.53 & 1.86$\pm$1.16 & 2.1$\pm$1.35 \\
    d) social and friendly & 6.77$\pm$0.50 & 4.06$\pm$1.43 & -&- & -& 6.36$\pm$1.18\\
  \end{tabular}
  \caption{Results of graphical affects validation survey. For certain clips, certain questions are irrelevant to the affect of interest, so we didn't ask those questions and left the table blank. The scores indicate the mean$\pm$standard deviation of the Likert scale (from 1 to 7) rating for the corresponding affect. The higher the score is, the happier/less calm/more energetic/having slept better the agent appears to the user. The first column describes the animated affect in the four clips.}~\label{tab:table1}
\end{table*}

We deploy various camera shots by scripting the Camera game object in Unity. Four rotating camera angles are used to reveal progress and the corgi's intelligence and emotions: a full frontal shot, a shot from the corgi's perspective, a close-up and a side shot. We also created seven special camera shots that are more relevant for certain subconscious psychological effects. For example, a Dutch Angle camera (tilted frame so subject is no longer parallel to the ground), frequently used in films to represent a disorientation, is used in the sleep event to show the slow loss of consciousness and state of fatigue or drowsiness. Likewise, in depicting the solitude of the corgi, there is a camera that slowly rotates above the corgi; here the small distance serves to detach the viewers from the subject and emphasize his smallness and loneliness. Several other special shots, including extreme close ups, panning of the weather, and more, can also be called when deemed appropriate. Lens effects include depth-of-field blur, recoloring, double focus, and so on. The result is a subconscious influence that subtly controls how the audience is perceiving each scene \cite{psychology2011lens,storytelling2017lens}. 

\section{Graphical affects validation study}
To examine the affective influence of the animation on the users, we divided the experiments into two steps. In the first step, we showed users short animated clips and validated that they perceive the desired emotions from our animation and cinematographic design. We can't possibly test all the possible scenes as a result of so many degrees of freedom in each scene, so we chose four clips that represent four basic affective states. We call this step "graphical affects validation". In the second step, we design a user study to examine the effect of the personalized animated movies.

In the graphical affects validation study, we conduct an online survey of 30 subjects to evaluate the emotional effects of four key animation clips, each showing the virtual avatar being a) stressed and anxious, b) happy and energetic, c) frustrated and tired and d) sociable and friendly. For example, in the video clip a), in order to indicate the high degree of stress and anxiety, the lighting in the forest is dark, the landscape morphs into obstacles, the background music is eerie, and the agent is staring at the forest, panting heavily. 

The user is then asked to rate corgi's happiness level, energy level, stress level, calmness level on a 1-7 Likert scale, and asked whether they have an experience they can relate to the video.

\textbf{Results of graphical affects validation.} The results of the graphical affects validation survey are shown in Table \ref{tab:table1}. The results overall confirmed that the intended states were perceived from the animations: On average, video clips b) and d) have a high happiness score (6.23 and 6.77),  while c) has a low happiness score (1.7). Users are generally able to perceive clip a) as a stressful situation (5.33), the dog in clip c) as sleepy or tired (1.86), and the dog in clip d) as sociable (6.36). The T-test to compare the mean scores of certain perceived emotions show statistical significance with users considering the happiness level in clip d) higher than c) by 3 scales (p<0.001), the happiness level in clip b) higher than a) by 1 scale (p=0.0019); the stress level in clip a) higher than b) by 2 scales (p<0.001); both the energetic level and sleep quality of the agent in video clip b) are 3-scales higher than c) (p=<0.001, p=0.0029).

The study also revealed some specific instances of ambiguity in the animations.  The responses of 8 out of 30 users indicated that it was unclear whether they thought clip a) reflects "excitement" or "anxiety" of the dog, while they mostly agreed the clip depicts "a stressful situation". It suggested that these users either perceive the stressful situation as a challenge (exited response) or threat (anxious response). Therefore, we incorporated two questions "how stressful was today" and "how did you handle the stress" in the survey and created different animations to separate these effects. (See Section \ref{section:stress} From Data to Animation - Stress).





Based on the promising results of graphical affects validation, we were able to create personalized animated movies with clips like these as building stones.

\section{A trip to the moon: Video study}
We conducted a one-week video study to evaluate the motivational effect of the personalized animated movies. The study protocol and recruitment process are pre-approved by the MIT Committee On the Use of Humans as Experimental Subjects (COUHES). 

We sent emails to university labs and dorms, recruiting 27 participants aged 18-36. Random assignment was made resulting in 13 of them assigned to the control group, and 14 to the personalized group. One participant in the control group dropped out of the study and asked to have his/her data deleted. All participants took pre-study surveys including tests on personality, stress level and mental health status. Each day during the week, participants reported their daily mood and behavioral data through an online Google Forms survey.

At the end of the week, participants received an animated movie through email, and were instructed to watch it. Members of the personalized group each received a story that was personalized according to each individual's mood and behavioral data, while the control group receive a non-personalized animated video of the same length, featuring a corgi dog rendered from the same models. Considering that the one-line cues are part of the personalization package, they were only used in the personalized group. The control video is generated from real data of a preliminary study participant, showing the corgi dog running through a forest, facing a stressful rain of rocks, having one negative and one positive interaction with another dog and in the end running uphill on a good note. The control video was picked such that it features some ups and downs that all participants might relate to their surveyed experience.

All participants were told the same story:  that the video was customized to their mood and behavior data. After watching the videos, participants filled out an evaluation form. There were three questions that accepted free text-form responses, without prompting the participants with any of the researcher's preconceptions such as "stress" or "social interaction":

\begin{itemize}
\item Q1 Ignoring imperfect AI renderings, what do you think about the story in the video?
\item Q2 What do you think about the main character (corgi dog)?
\item Q3 This video is generated from your mood and behavioral data. What do you think the video reflected about you?
\end{itemize}

\section{FINDINGS}

After one week, we received 10 responses to the video through Google Forms from the 12 people remaining active in the control group and 13 responses to the video from the 14 people in the personalized group. We first present a quantitative overview of the results by looking at behavioral measures such as the lengths of their responses, and then do a thematic analysis \cite{thematic2006} to discuss participants' emotional engagement, connection to the agent and change in self-reflection.

Two researchers first read the responses with the research questions in mind, blinded to conditions, and came up with four quantitative cross-group comparisons. One researcher then coded the responses manually, blinded to conditions. First, we looked at how many participants in each group showed confusion about the story, by saying "I was confused", "not clear what is happening". Second, we compared the average length of response in two groups. Third, we counted how many emotion-descriptive (e.g. "emotional", "nostalgic") words they used to describe themselves. Fourth, we compared if they recalled a past experience from the animation. The comparison results are listed in Table \ref{tab:groupdifference}. The results speak positively about the engaging effect of personalized animations (less confusing, more emotionally engaging, inducing lengthier writings and more recalling of past experiences), but are not statistically significant.

Considering that simple quantitative measures as above don't capture the full picture, we read the responses and found that participants' responses fell into three types: indifferent, intrigued, affected, based on the length of their response and whether they connected the animation to themselves.

\textbf{Indifferent.} Participants who were indifferent showed confusion and negativity when asked what they think of the story. To Q1, they replied with confusion. To Q3, they denied that the video represents or reflects them and didn't relate the video to their personal experience. Their answers are all brief, adding up to fewer than around 80 words (M=42.8, SD=27.1), showing a low level of engagement.

\textbf{Intrigued.} Participants who were intrigued provided their interpretation of the story to Q1. They had some connection with the agent by describing the corgi dog with positive words such as "cute", "expressive". They wrote short but positive answers, the total word count was fewer than around 80 (M=40.1, SD=26.6). They were able to discover one or two self-reflections.

\textbf{Invested.} Participants who were invested used the most emotion-related words to describe their feelings for the video. They wrote the longest answers, ranging from around 80 to over 300 words (M=127, SD=93.9, the big standard deviation due to the long tail of long responses). They described a strong personal connection with the agent and could describe multiple reflections on their experiences and personalities. The defining factors of the three types of response are summarized in Table \ref{tab:table2}.

\begin{table}[h!]
  \centering
  \begin{tabular}{l|r r r r}
    {\small\textit{Group}}
    & {\small \textit{Confused}}
      & {\small \textit{Word Count}}
    & {\small \textit{Emotion}}
    & {\small \textit{Recall}}
    \\
    \midrule
    Control & 50\% & 71.3 $\pm$ 45.7 & 2.2 $\pm$ 2.5  &60\% \\
    Personalized & 30.8\% & 84.5 $\pm$ 95.4 & 3.3 $\pm$ 3.0 &84.6\% \\  
   	P-values & All > 0.1 & & &\\
  \end{tabular}
  \caption{Comparison between the control group and the personalized group. Confused refers to the percentage of users showing confusion about the story (e.g. By saying "Don't know what's going on", "I am confused"). Word count is the group average of word count. Emotion refers to the average number of emotion-descriptive words (e.g. "emotional", "happy", "nostalgic") the user used in their response. Recall refers to the percentage of participants recalling past experiences corresponding to the animation in the group.}~\label{tab:groupdifference}
\end{table}

\begin{table}[h!]
  \centering
  \begin{tabular}{p{4cm}|r r r}
    {\small\textit{Type of response}}
    & {\small \textit{Indifferent}}
      & {\small \textit{Intrigued}}
    & {\small \textit{Invested}}
    \\
    \midrule
    Word count & < 80 & < 80 & > 80 \\
    Confusion & Yes & No & No \\
    Described emotions they felt & No & Yes & Yes \\
    Wrote self-reflection & No & Yes & Yes \\
  \end{tabular}
  \caption{Three types of responses to the video.}~\label{tab:table2}
\end{table}

\begin{figure}[h!]
\centering
  \includegraphics[width=0.9\columnwidth]{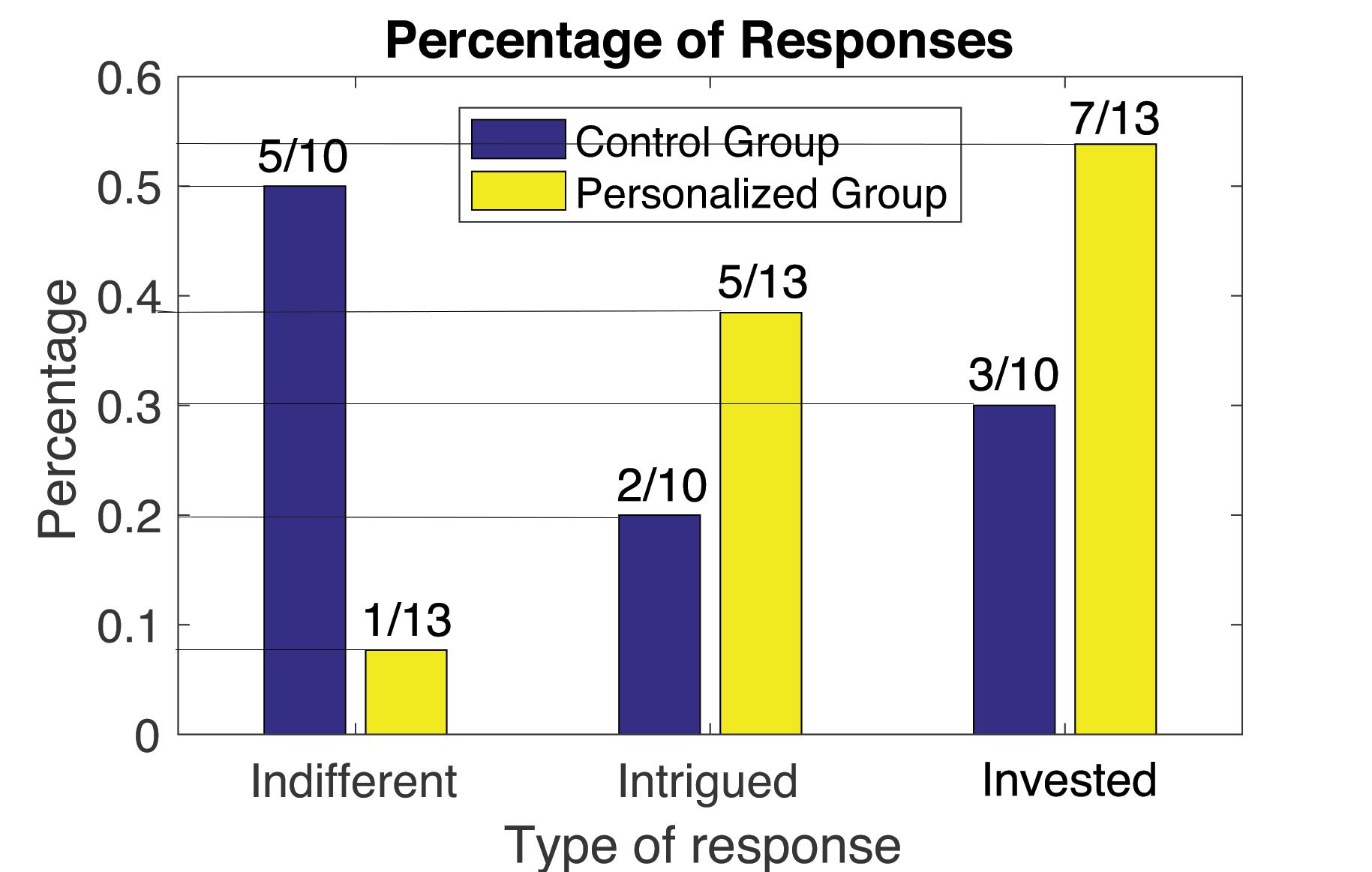}
  \caption{Distribution of three types of responses among the two study groups.}
  \label{fig:responses}
\end{figure}

The distribution of the three types of responses in the two study groups is shown in Fig. \ref{fig:responses}. While the overall study is limited to a relatively small number of people, we can see higher intrigued and invested frequencies in the personalized group than in the control group. Half (5 out of 10) participants in the control group showed indifference about the video, while only 1 out of 13 participants in the personalized group showed indifference. The difference in  types of responses in the two groups is statistically significant (p=0.036 at confidence level 0.95) using an independent-samples unequal-variance t-test. The result suggests that while participants in both groups try to reflect on their behavior when told the video is generated from their personal data, the personalized videos were more interpretable and successful in terms of eliciting personal reflection than was a generic, non-personalized animation (even though the participants in both groups were told it was "personalized".)


\subsection{Emotional engagement, or "beautifully poignant."}

Below demonstrate the effect of personalized animation with examples. We mostly list responses from the personalized group, omitting those of the control group, half of which describe confusion and indifference anyway. Interested readers can refer to \href{https://docs.google.com/spreadsheets/d/1jadgvy0SQ6TZYjDExqMZTYwRpE4oUSwMyBP9gT2T2U4/edit?usp=sharing}{this online database} for a full collection of participant responses.

Among the intrigued and invested participants, emotion responses range from low arousal ("calm") to high arousal ("super excited"), low valence ("sad", "didn't like") to high valence ("glad"), and more complex emotions such as nostalgia and independence. Below are some answers to Q1, "what do you think about the story", from participants whose age, gender and group (P for personalized, C for control) are included in brackets.

Happy: \textit{"I feel glad that the corgi was able to walk on the moon after a few chapters featuring ups and downs."} (25, male, P)

Sad: \textit{"It made me feel nostalgic and at some points a little sad."} (18, male, P) \textit{"It seemed kind of aimless and repetitive at times, but was also beautifully poignant at others."} (19, male, P)

Calm: \textit{"I feel calm and relaxed."} (30, male, P)

Unsettling: \textit{"It was scary and unsettling in the middle, but the nice music as I/the dog chilled on the moon made it slightly better."} (19, female, P)

Dynamic and evolving: \textit{"The very beginning made me super excited, because corgis are great and I recognized the phrases I entered. I could tell the story began happily. The video certainly dipped into a spooky atmosphere, and I felt anxious watching the corgi become separated from its friend by a wall of rocks. The middle made me nervous that something was going to pop out on the screen. The end greatly confused me, and I felt neither happy nor sad. But it seemed the corgi wanted to reach the moon all along, and it finally achieved its goal."} (20, female, P)

\textbf{Denial of one's behavior.} Some participants were engaged by the story, but showed a negative taste for certain behaviors of the agent. \textit{"I didn't like the story. The corgi didn't really interact with the other dog and was alone at the end."} (19, male, P) 

This attitude of denial indicates that the participant might have felt self-conscious or offended when seeing their undesired behaviors animated on a character. The same participant responded, \textit{"I have a hard time connecting with others,"} when asked about his self-reflection in Q3. This response is consistent with the behavior he disliked of the agent. It suggests that seeing a truthful reflection of one's undesired behavior could be a relatively unpleasant emotional experience. 

From this exploration, and the seriousness with which participants reflected based on this automated animated portrayal of their data, we believe that personalized story videos should, carefully balance the unpleasantness with hope and positivity.  We do not wish to have videos presenting bad mood or unwanted behavior data leave a participant unable to overcome the initial discomfort; rather, we want to enable them to emotionally reflect, but then also to feel motivated to change their behavior. As another participant pointed out, \textit{"I was kind of shocked to see things suddenly get dark, but I think it also made sense... It was just a little jarring to come face to face with."} (19, female, P).

Notice that certain participants in the control group (5/10) were also emotionally invested in the video. One 
participant wrote \textit{"It left me with this feeling like I can be independent which is something I'm also working on after years of being dependent on significant others and because I'm still recovering from a hard break-up, seeing the corgi realize he didn't (rely on) the other dog to be happy made me happy."} (22, female, C) We consider such high engagement with the video to be a positive sign that our animation and cinematographic design were successful in connecting emotionally with viewers.

\subsection{Connection to the agent, or "I/the corgi"}
We analyzed the participants' empathy for the agent by looking at replies to Q3, "What do you think about the corgi dog?"

Of the 13 participants in the personalized group, 8 directly described their connection to the dog. \textit{"The dog seemed kind of lonely for most of the video. For some reason (maybe its facial expression and tongue flapping) the dog appeared naive and kind of stupid to me, but also always hopeful. I felt a deep emotional connection to the dog at times."} (19, male, P) \textit{"I like it, it is cute and expressive."} (18, male, P) \textit{"The corgi is adorable. I was sad when the dog looked sad/sleepy. While watching it, I thought of how I could make the corgi happier and then at the end, it was so happy :D"} (22, female, P) One participant showed that they identified with the agent by referring to them and the agent as the same entity: \textit{"I/the dog chilled on the Moon"}.

Some participants' attitude towards the agent was more descriptive of what they saw it looked like and less descriptive of their personal connection. \textit{"His mood changes very quickly."} (19, male, P) \textit{"Temperamental."} (21, female, P) An 18-year-old female participant (P) considered the dog \textit{"high"}, which was not one of the intended states, although it could be possible to also ask participants about their drug and alcohol behavior and reflect such use on their avatar. A 19-year-old male participant (C) said the agent \textit{"has communication and social intelligence problems"}. His reply to Q3, \textit{"That I was confused and in a mentally bad place,"} seems to indicate that he may have empathized with the agent's visually-demonstrated challenges and problems.

\subsection{Self-reflection, or "needs more friends"}

The one-line reminders provided daily by each participant appeared as chapter titles in their personalized animations, which served as memory retrieval cues. Two participants explicitly commented on the effect of the retrieval cues, \textit{"I was also vaguely amazed at how well I could recall each day based on the words in the chapter names <- which also kept me invested, because I knew it was me."} (19, female, P) \textit{"The very beginning made me super excited, because corgis are great and I recognized the phrases I entered."} (20, female, P)

Participants independently came to conclusions about their mood behavior pattern when answering Q3 "what do you think the video reflected about you". Notice that they were not prompted to write about "friends", "exercise" or "stress", nor was there any text in the animation indicating that a particular scene was about a certain behavior. In other words, the reflections were solely based on their understanding of the story world and memory of doing the surveys.

\textbf{Overall mood.} 
\textit{"It reflected that I have been fairly happy the past couple of days."} (30, male, P)

\textit{"I've been through a lot recently, but try to keep my head up and can do so because of my friends."} (21, female, C)

\textit{"Mostly happy with some occasional stress, maybe the interaction with the other dog represents an interaction with another person, also I sometimes think of my personality as a geometrically rendered dog."} (20, female, C)

\textbf{Social life.} 
The animated video portrayed the self-reported social interactions as interactions between their avatar and another dog, prompting some to reflect on their social life and its interaction with their mood.  A number of participants reflected in writing on aspects of their social life that need improvement, such as loneliness and isolation. \textit{"I have a hard time connection with others."} (19, male, P) \textit{"Needs more friends."} (18, female, P) \textit{"When I isolate myself, I tend to be sad and unproductive. I sometimes need time away from people, but that can turn into a negative thing when there's too much of it."} (19, female, C) \textit{"The presence of other dogs seems to reflect how my mood strongly depends on my interactions with other people. When the corgi got walled off by all those rocks, I was reminded how lonely I feel when I go without human interaction for too long."} (20, female, P)

\textbf{Stress.} When there was a stress scene in the personalized videos, participants were usually able to identify the tense atmosphere and connect it with their stressful experience. \textit{"The wall of rocks also corresponded well with all the homework I received this past week."} (20, female, P) \textit{"That I am sometimes calm and collected but other times I am overwhelmed by stress and wants to run away/not know how to handle the stress."} (22, female, P) One participant (P) was, however, unable to identify the source of the stress, \textit{"(The video reflects) that I'm angry?? It made the world seem very scary and out to get me. Why was this dog on a planet in the middle of nowhere with hostile boulders and angry other dogs?"}

\textbf{Sleep and exercise.} Fewer participants discovered the relationship between sleep, exercise, and mood. One 18-year-old male participant (P) thought he could have been more "active", which could either refer to a mental or physical state. No participant mentioned "sleep", though some of them had very irregular sleeping schedules, causing the agent to appear tired and fall asleep often in the video.

It is interesting how certain participants also came to a conclusion about mental states that are not included in the surveys, and thus not intentionally included in the animations. In other words, some thought that the animated video, or the system behind it, had the intelligence to speculate on their "hidden states:"  

\textit{"The video also implies I'm more of a follower, since I'm not the type to push people to do things. I'm not sure what the moon represents, but if the video is trying to imply I space out a lot, it'd be very correct."} (20, female, P)

\textit{"I've mostly been around people this week, but sometimes, I do feel a little lonely off and on; I'm surprised the video caught that even though there's no question asking if I feel lonely, so I'm impressed that this "AI" somehow captured that vulnerability in me."} (22, female, C)

\section{Discussion}
\textbf{Focus on behaviors of interest.}
In order to better represent each user's past week, also in response to users' request in the preliminary study, our video study involved mood and four types of behaviors. 
As a result, it was harder for participants to intuitively understand how their sleep and exercise affected the story. Future videos can try to focus on fewer dimensions of data and dedicate the story to emphasize the largest change in the data, or a change that the user specifies is of particular interest to them. Slight changes to the automation algorithm could enable their personal animation to function as a kind of amplifier of the behaviors on which they most want to reflect.

\textbf{Balance hope and negativity.}
Emotionally engaging stories require the alternation of ups and downs in the agent's experiences, and can either have a happy ending or a sad ending. However, for some audiences, the negative part of the story is more difficult to come face to face with. Certain participants found the dark part of the story "jarring", "scary" and "unsettling". Our study findings suggest that while negativity might seem discouraging to some individuals, it can be carefully balanced with positivity and the hope to change. Future work needs to consider how to handle this balance when all the data from the participant is negative.

As movie creators, we are not constrained by any set of rules to present the emotions of the movie. As a result, we have the ability to "manipulate" users' perception of their behaviors. We can zoom in on a positive experience or dwell on a negative one, depending on the desired cognitive and psychological response. Future studies can dig more into these effects on a personalized basis. 

\textbf{Judging good work}
The above discussion naturally leads to the question: what do we define as "good" personalized animation? From our interaction with participants, we see personalized animation potentially serving different purposes: self-reflection, changing behavior, or even communication of emotions. Depending on the purpose, the standards for "good personalized animation" can vary. Generally, a good personalized video should be able to elicit curiosity and empathy - viewers keep anticipating what happens next out of compassion for the characters (themselves); subsequently, what designers want users to do with that feeling can be tailored from case to case.

\textbf{Limitations.}
The results here should be taken with some caution, as our sample size was small and the study was conducted only for one week for each of the 27 participants.  Our next steps will involve longer studies where we validate participants' behavior change in the long term (usually months). The current system can support an emotionally-engaging, 6-8 minute animation for one individual. However, will the believability of the story decrease when users see repetitive content, especially when some people have routines so that their weekly data is really similar? To increase believability, we currently try to randomize the agent's path on the landscape. In the future we will also enlarge the animation and model database to provide more visible dynamics.

From our post-study survey of 27 participants, the average interest in watching more videos from future data is 5.25 on a Likert scale of 1 to 7, indicating that they are curious overall to see changes in their personal animation story when their behaviors and experiences change.
Moving forward, we will investigate emotional engagement and behavior change induced by personalized animated movies delivered periodically (such as weekly or monthly). We will look at their interactions with the animations and how impact on their feelings changes over time.
We will also try to measure emotional engagement by more objective signals than the current self-reported data, including facial expressions, and sensor-based measures such as electrodermal responsivity. Behavioral measures such as whether participants have shared their video with friends and family can also be used. Objective signals measured over the course of a day, including physical activity and sleep activity as well, can be combined with written responses to provide a more complete affective-cognitive-physical state assessment.

\section{Conclusion}
Converting personal data into a format that can have emotional impact can potentially fuel motivation for self-reflection and positive behavior change; however, a fully-automated system is not capable today of understanding human personal experiences at the kind of level that could make such a story both explicit and accurate.  To solve this challenge, we built an automated system to construct an animation that is personalized to the data provided by an individual over a week, utilizing an animated avatar (corgi) that implicitly reflects the person's behaviors, such as sleep and social interactions, and moods associated with each day of the week.  We conducted two studies: The first found that the avatar's portrayal of a set of moods and behaviors could be perceived accurately in general. The second study tested the emotional engagement of using a personalized story against a challenging control: a similar story that they were \textit{told} was personalized. The results indicated that personalized animations tended to be more emotionally engaging, encouraging greater and lengthier writing that indicated self-reflection about moods and behaviors.  Furthermore, while human imagination plays an important and valuable part in both conditions -- test and control -- the impact we found suggests that true personalization may be more powerfully influential on moods and self-reflection than simply believing that one is receiving personalized feedback.

\section{Acknowledgement}
This research was partially supported by the Advancing Wellbeing Initiative at the MIT Media Lab. We thank Asma Ghandeharioun, Weixuan (Vincent) Chen and Sam Spaulding for comments that greatly improved the manuscript, and 3 "anonymous" reviewers for their insights in presenting the research methodology and results.

\bibliographystyle{SIGCHI-Reference-Format}
\bibliography{sample}


\begin{thebibliography}{00}


\ifx \showCODEN    \undefined \def \showCODEN     #1{\unskip}     \fi
\ifx \showDOI      \undefined \def \showDOI       #1{{\tt DOI:}\penalty0{#1}\ }
  \fi
\ifx \showISBNx    \undefined \def \showISBNx     #1{\unskip}     \fi
\ifx \showISBNxiii \undefined \def \showISBNxiii  #1{\unskip}     \fi
\ifx \showISSN     \undefined \def \showISSN      #1{\unskip}     \fi
\ifx \showLCCN     \undefined \def \showLCCN      #1{\unskip}     \fi
\ifx \shownote     \undefined \def \shownote      #1{#1}          \fi
\ifx \showarticletitle \undefined \def \showarticletitle #1{#1}   \fi
\ifx \showURL      \undefined \def \showURL       #1{#1}          \fi

\bibitem{baer2006using}
{Ruth~A Baer}, {Gregory~T Smith}, {Jaclyn Hopkins}, {Jennifer Krietemeyer},
  {and} {Leslie Toney}. 2006.
\newblock \showarticletitle{Using self-report assessment methods to explore
  facets of mindfulness}.
\newblock {\em Assessment\/} {13}, 1 (2006), 27--45.
\newblock


\bibitem{bates1994role}
{Joseph Bates} {and} {others}. 1994.
\newblock \showarticletitle{The role of emotion in believable agents}.
\newblock {\it Commun. ACM} {37}, 7 (1994), 122--125.
\newblock


\bibitem{bentley2005evaluation}
{Todd Bentley}, {Lorraine Johnston}, {and} {Karola von Baggo}. 2005.
\newblock \showarticletitle{Evaluation using cued-recall debrief to elicit
  information about a user's affective experiences}. In {\em Proceedings of the
  17th Australia conference on Computer-Human Interaction: Citizens Online:
  Considerations for Today and the Future}. Computer-Human Interaction Special
  Interest Group (CHISIG) of Australia, 1--10.
\newblock


\bibitem{thematic2006}
{Virginia Braun} {and} {Victoria Clarke}. 2006.
\newblock \showarticletitle{Using Thematic Analysis in Psychology}.
\newblock   {3} (01 2006), 77--101.
\newblock


\bibitem{costa2016emotioncheck}
{Jean Costa}, {Alexander~T Adams}, {Malte~F Jung}, {Fran{\c{c}}ois
  Guimbetiere}, {and} {Tanzeem Choudhury}. 2016.
\newblock \showarticletitle{EmotionCheck: leveraging bodily signals and false
  feedback to regulate our emotions}. In {\em Proceedings of the 2016 ACM
  International Joint Conference on Pervasive and Ubiquitous Computing}. ACM,
  758--769.
\newblock


\bibitem{freytag1896freytag}
{Gustav Freytag}. 1896.
\newblock {\em Freytag's technique of the drama: an exposition of dramatic
  composition and art}.
\newblock Scholarly Press.
\newblock


\bibitem{gobel2010serious}
{Stefan G{\"o}bel}, {Sandro Hardy}, {Viktor Wendel}, {Florian Mehm}, {and}
  {Ralf Steinmetz}. 2010.
\newblock \showarticletitle{Serious games for health: personalized exergames}.
  In {\em Proceedings of the 18th ACM international conference on Multimedia}.
  ACM, 1663--1666.
\newblock


\bibitem{gustafsson2009power}
{Anton Gustafsson}, {Magnus B{\aa}ng}, {and} {Mattias Svahn}. 2009.
\newblock \showarticletitle{Power explorer: a casual game style for encouraging
  long term behavior change among teenagers}. In {\em Proceedings of the
  International Conference on Advances in Computer Enterntainment Technology}.
  ACM, 182--189.
\newblock


\bibitem{harlow1986depression}
{Lisa~L Harlow}, {Michael~D Newcomb}, {and} {Peter~M Bentler}. 1986.
\newblock \showarticletitle{Depression, self-derogation, substance use, and
  suicide ideation: Lack of purpose in life as a mediational factor}.
\newblock {\em Journal of clinical psychology\/} {42}, 1 (1986), 5--21.
\newblock


\bibitem{hook2009affective}
{Kristina H{\"o}{\"o}k}. 2009.
\newblock \showarticletitle{Affective loop experiences: designing for
  interactional embodiment}.
\newblock {\em Philosophical Transactions of the Royal Society of London B:
  Biological Sciences\/} {364}, 1535 (2009), 3585--3595.
\newblock


\bibitem{isbister2007sensual}
{Katherine Isbister}, {Kia H{\"o}{\"o}k}, {Jarmo Laaksolahti}, {and} {Michael
  Sharp}. 2007.
\newblock \showarticletitle{The sensual evaluation instrument: Developing a
  trans-cultural self-report measure of affect}.
\newblock {\em International journal of human-computer studies\/} {65}, 4
  (2007), 315--328.
\newblock


\bibitem{kihlstrom2000emotion}
{John~F Kihlstrom}, {Eric Eich}, {Deborah Sandbrand}, {and} {Betsy~A Tobias}.
  2000.
\newblock \showarticletitle{Emotion and memory: Implications for self-report}.
\newblock {\em The science of self-report: Implications for research and
  practice\/} (2000), 81--99.
\newblock


\bibitem{psychology2011lens}
{Kurt Lancaster}. 2011.
\newblock \showarticletitle{The Psychology of the Lens: Patrick Moreau creates
  filmic intimacy with DSLRs at Stillmotion}.
\newblock  (June 2011).
\newblock


\bibitem{storytelling2017lens}
{Kurt Lancaster}. 2017.
\newblock \showarticletitle{Using lenses to enhance visual storytelling}.
\newblock  (January 2017).
\newblock


\bibitem{lindstrom2006affective}
{Madelene Lindstr{\"o}m}, {Anna St{\aa}hl}, {Kristina H{\"o}{\"o}k}, {Petra
  Sundstr{\"o}m}, {Jarmo Laaksolathi}, {Marco Combetto}, {Alex Taylor}, {and}
  {Roberto Bresin}. 2006.
\newblock \showarticletitle{Affective diary: designing for bodily
  expressiveness and self-reflection}. In {\em CHI'06 extended abstracts on
  Human factors in computing systems}. ACM, 1037--1042.
\newblock


\bibitem{mateas2003narrative}
{Michael Mateas} {and} {Phoebe Sengers}. 2003.
\newblock Narrative intelligence.
\newblock   (2003).
\newblock


\bibitem{mccrae1984situational}
{Robert~R McCrae}. 1984.
\newblock \showarticletitle{Situational determinants of coping responses: Loss,
  threat, and challenge.}
\newblock {\em Journal of personality and Social Psychology\/} {46}, 4 (1984),
  919.
\newblock


\bibitem{michael2005serious}
{David~R Michael} {and} {Sandra~L Chen}. 2005.
\newblock {\em Serious games: Games that educate, train, and inform}.
\newblock Muska \& Lipman/Premier-Trade.
\newblock


\bibitem{picard2005evaluating}
{Rosalind~W Picard} {and} {Shaundra~Bryant Daily}. 2005.
\newblock \showarticletitle{Evaluating affective interactions: Alternatives to
  asking what users feel}. In {\em CHI Workshop on Evaluating Affective
  Interfaces: Innovative Approaches}. ACM New York, NY, 2119--2122.
\newblock


\bibitem{picard1997affective}
{Rosalind~W Picard} {and} {Roalind Picard}. 1997.
\newblock {\em Affective computing}. Vol. 252.
\newblock MIT press Cambridge.
\newblock


\bibitem{porter2000site}
{Tom Porter} {and} {Galyn Susman}. 2000.
\newblock \showarticletitle{On site: Creating lifelike characters in pixar
  movies}.
\newblock {\it Commun. ACM} {43}, 1 (2000), 25.
\newblock


\bibitem{rosling2006debunking}
{Hans Rosling}. 2006.
\newblock {\em Debunking third-world myths with the best stats you've ever
  seen}.
\newblock TED.
\newblock


\bibitem{roth2009motivational}
{Christian Roth}, {Peter Vorderer}, {and} {Christoph Klimmt}. 2009.
\newblock \showarticletitle{The motivational appeal of interactive
  storytelling: Towards a dimensional model of the user experience}. In {\em
  Joint International Conference on Interactive Digital Storytelling}.
  Springer, 38--43.
\newblock


\bibitem{sano2016measuring}
{Akane Sano}. 2016.
\newblock {\em Measuring college students' sleep, stress, mental health and
  wellbeing with wearable sensors and mobile phones}.
\newblock Ph.D. Dissertation. Massachusetts Institute of Technology.
\newblock


\bibitem{sas2013affectcam}
{Corina Sas}, {Tomasz Fratczak}, {Matthew Rees}, {Hans Gellersen}, {Vaiva
  Kalnikaite}, {Alina Coman}, {and} {Kristina H{\"o}{\"o}k}. 2013.
\newblock \showarticletitle{AffectCam: arousal-augmented sensecam for richer
  recall of episodic memories}. In {\em CHI'13 Extended Abstracts on Human
  Factors in Computing Systems}. ACM, 1041--1046.
\newblock


\bibitem{sengers2005reflective}
{Phoebe Sengers}, {Kirsten Boehner}, {Shay David}, {and} {Joseph'Jofish' Kaye}.
  2005.
\newblock \showarticletitle{Reflective design}. In {\em Proceedings of the 4th
  decennial conference on Critical computing: between sense and sensibility}.
  ACM, 49--58.
\newblock


\bibitem{sengers2006staying}
{Phoebe Sengers} {and} {Bill Gaver}. 2006.
\newblock \showarticletitle{Staying open to interpretation: engaging multiple
  meanings in design and evaluation}. In {\em Proceedings of the 6th conference
  on Designing Interactive systems}. ACM, 99--108.
\newblock


\bibitem{snyder1977acceptance}
{CR Snyder}, {Randee~J Shenkel}, {and} {Carol~R Lowery}. 1977.
\newblock \showarticletitle{Acceptance of personality interpretations: the"
  Barnum Effect" and beyond.}
\newblock {\em Journal of consulting and clinical psychology\/} {45}, 1 (1977),
  104.
\newblock


\bibitem{thomas1995illusion}
{Frank Thomas}, {Ollie Johnston}, {and} {Frank. Thomas}. 1995.
\newblock {\em The illusion of life: Disney animation}.
\newblock Hyperion New York.
\newblock


\bibitem{van2004bringing}
{AJN Van~Breemen}. 2004.
\newblock \showarticletitle{Bringing robots to life: Applying principles of
  animation to robots}. In {\em Proceedings of Shapping Human-Robot Interaction
  workshop held at CHI 2004}. 143--144.
\newblock


\end{thebibliography}

\end{document}